\documentclass[12pt,pre,aps,showpacs,preprint,nofootinbib]{revtex4}

\usepackage{epsfig}
\usepackage{graphicx}
\usepackage{amsfonts}
\usepackage{amssymb}
\usepackage{amsmath}

\newcommand{\beq}{\begin{equation}}
\newcommand{\eeq}{\end{equation}}
\newcommand{\avg}[1]{\left< #1 \right>}

\newcommand{\barr}{\begin{eqnarray}}
\newcommand{\earr}{\end{eqnarray}}
\newcommand{\Ord}[1]{{\cal O}\left( #1\right)}

\begin{document}

\title{Estimates of the optimal density and kissing number of sphere packings in high
dimensions}

\author{A. Scardicchio}

\email{ascardic@princeton.edu}

\affiliation{\emph{Department of Physics}, \emph{Joseph Henry Laboratories},\emph{Princeton University}, Princeton
NJ 08544}

\affiliation{\emph{Princeton Center for Theoretical Physics, Princeton University}, Princeton
NJ 08544}

\author{F. H. Stillinger}

\email{fhs@princeon.edu}

\affiliation{\emph{Department of Chemistry}, \emph{Princeton University}, Princeton
NJ 08544}

\author{S. Torquato}

\email{torquato@electron.princeton.edu}

\affiliation{\emph{Department of Chemistry}, \emph{Princeton University}, Princeton
NJ 08544}

\affiliation{\emph{Program in Applied and Computational Mathematics}, \emph{Princeton
University}, Princeton NJ 08544}

\affiliation{\emph{PRISM, Princeton University}, Princeton NJ 08544}

\affiliation{\emph{Princeton Center for Theoretical Physics, Princeton University}, Princeton
NJ 08544}

\begin{abstract}

The problem of finding the asymptotic behavior of the maximal density $\phi_{\mbox{\scriptsize max}}$
of sphere packings in high Euclidean dimensions is one of the most fascinating and challenging problems in 
discrete geometry. One century ago, Minkowski obtained a rigorous lower
bound on $\phi_{\mbox{\scriptsize max}}$ that is controlled asymptotically by $1/2^d$, where
$d$ is the Euclidean space dimension. An indication of the difficulty of the problem can be 
garnered from the fact that  exponential improvement of Minkowski's bound has proved to be elusive, 
even though existing upper bounds suggest that such improvement should be possible.
Using a statistical-mechanical procedure to optimize the density associated with a ``test" pair correlation 
function and a conjecture concerning the existence of disordered sphere packings
[S. Torquato and F. H. Stillinger, Experimental Math. {\bf 15}, 307 (2006)], the putative exponential
improvement on $\phi_{\mbox{\scriptsize max}}$
was found with an asymptotic behavior controlled by $1/2^{(0.77865\ldots)d}$.
Using the same methods,  we investigate whether this exponential improvement can be further improved 
by exploring other test pair correlation functions correponding
to disordered packings. We demonstrate that there are simpler test functions that lead to
the same asymptotic result. More importantly, we show that there is a wide class of test functions that lead
to precisely the same putative exponential  improvement and therefore
the asymptotic form $1/2^{(0.77865\ldots)d}$ is much more general than  previously surmised.
This class of test functions leads to an optimized average kissing number 
that is controlled by the same asymptotic behavior as the one found in the aforementioned paper.

\end{abstract}

\maketitle

\section{Introduction}

A collection of congruent spheres in $d$-dimensional Euclidean space $\mathbb{R}^d$
is called a {\it sphere packing} if no two spheres overlap.
Although the practical relevance of sphere packings in high Euclidean dimensions 
was appreciated by Shannon in 1948 \cite{Sh48}, there has been a resurgence
of interest in such problems in both the physical and mathematical sciences \cite{To02c,
Co02,Co03,Ha05,To06a,To06b,Pa06,Sk06,To06c,Co07}. Shannon showed
that the optimal way of sending digital signals over noisy channels corresponds
to the densest sphere packing in a high dimensional space \cite{Sh48}.
These ``error-correcting" codes underlie a variety of systems in digital
communications and storage \cite{Co93}.
Physicists have investigated sphere packings in high dimensions to gain insight
into classical ground and glassy states of matter as well as
phase behavior in lower dimensions \cite{Fr99,Pa00,To02a,Pa06,Sk06}.
Understanding the symmetries and other mathematical
properties  of the densest packings
in arbitrary dimension is a problem of long-standing interest in discrete geometry \cite{Co93,Ro58,Ro64,Co03,Ha05}.

The {\it packing density} or simply density $\phi$ of a sphere packing is the fraction of
space $\mathbb{R}^d$ covered by the spheres.
 We will call
\begin{equation}
\phi_{\mbox{\scriptsize max}}= \sup_{P\subset \mathbb{R}^d} \phi(P)
\end{equation}
the {\it maximal density}, where the supremum is taken over
all packings in $\mathbb{R}^d$. The set of lattice packings is a subset of the set of sphere
packings in $\mathbb{R}^d$. A {\it lattice} $\Lambda$ in $\mathbb{R}^d$ is a subgroup
consisting of the integer linear combinations of vectors that constitute a basis for $\mathbb{R}^d$.
A {\it lattice packing} $P_L$ is one in which  the centers of nonoverlapping spheres
are located at the points of $\Lambda$.
In a lattice packing, the space $\mathbb{R}^d$ can be geometrically divided into identical
regions $F$ called {\it fundamental cells}, each of which contains the center
of just one sphere. In the physical sciences, a lattice packing
is simply a packings arranged on the sites of a Bravais lattice.
Non-lattice packings include periodic packings (more than one
sphere per fundamental cell) as well as disordered
packings \cite{footnote1}.

The sphere packing problem seeks to answer the following  
question: Among all packings of congruent spheres,
what is the maximal packing density $\phi_{\mbox{\scriptsize max}}$, i.e., largest
fraction of $\mathbb{R}^d$ covered by the spheres,
and what are the corresponding arrangements of the spheres \cite{Ro64,Co93}?
For arbitrary $d$, the sphere packing problem  is notoriously difficult to solve.
In the case of packings of congruent $d$-dimensional
spheres, the exact solution is known for the first
three space dimensions. For $d=1$, the answer is trivial
because the spheres tile the space so that $\phi_{\mbox{\scriptsize max}}=1$.
In two dimensions, the optimal solution is the triangular lattice 
arrangement (also called the hexagonal packing) with $\phi_{\mbox{\scriptsize max}}=\pi/\sqrt{12}$.
In three dimensions, the Kepler conjecture that the face-centered cubic lattice arrangement
provides the densest packing with $\phi_{\mbox{\scriptsize max}}=\pi/\sqrt{18}$
was only recently proved by Hales \cite{Ha05}. For $3< d <10$,
the densest known packings of congruent spheres are lattice
packings (defined below). For example, the ``checkerboard" lattice $D_d$, which is
the $d$-dimensional generalization of the FCC lattice, is believed
to be optimal in $\mathbb{R}^4$ and $\mathbb{R}^5$. 
The $E_8$ and Leech lattices in $\mathbb{R}^8$
and $\mathbb{R}^{24}$, respectively,
are remarkable dense and symmetric and are most
likely the densest packings in these dimensions \cite{Co07}. However, for sufficiently large $d$,
lattice packings are most likely not the densest,
but it becomes increasingly difficult to find specific
dense packing constructions in high dimensions \cite{footnote2}.
In high dimensions, the best that one can do theoretically
is to devise rigorous upper and lower bounds on 
$\phi_{\mbox{\scriptsize max}}$.

Upper and lower bounds on the maximal density $\phi_{\mbox{\scriptsize max}}$ exist in all dimensions \cite{Co93}.
Minkowski \cite{Mi05} proved that the maximal density $\phi^L_{\mbox{\scriptsize max}}$ among all Bravais lattice packings
for $d \ge 2$ satisfies the lower bound
\begin{equation}
\phi^L_{\mbox{\scriptsize max}} \ge \frac{\zeta(d)}{2^{d-1}},
\label{mink}
\end{equation}
where $\zeta(d)=\sum_{k=1}^\infty k^{-d}$ is the Riemann zeta function.
One observes that for large values of $d$,
the asymptotic behavior of the {\it nonconstructive} Minkowski lower bound is controlled by $2^{-d}$.
Since 1905, many extensions and generalizations of (\ref{mink})
have been obtained \cite{Co93}, but none of these investigations have been able to improve
upon the dominant exponential term $2^{-d}$. The best currently known rigorous lower
bound on $\phi^L_{\mbox{\scriptsize max}}$ was obtained by Ball \cite{Ball92}.
He found that
\begin{equation}
\phi^L_{\mbox{\scriptsize max}} \ge \frac{2(d-1)\zeta(d)}{2^{d}}.
\label{ball}
\end{equation}
Interestingly, the density of a {\it saturated} packing of congruent spheres
in $\mathbb{R}^d$ for all $d$ satisfies the lower bound \cite{foot}
\begin{equation}
\phi \ge \frac{1}{2^d},
\label{sat}
\end{equation}
and thus  has  the same dominant exponential term as the Minkowski lower bound (\ref{mink}).
A {\it saturated packing} of congruent spheres
of unit diameter  and density $\phi$ in $\mathbb{R}^d$ has the property that each point in space lies
within a unit distance from the center of some sphere.
As we will discuss below, the lower bound (\ref{sat}) is not a stringent
bound for a saturated packing and therefore is improvable. 

Rogers \cite{Ro58,Ro64} found upper bounds on the
maximal density $\phi_{\mbox{\scriptsize max}}$
by an analysis of the Voronoi cells. For large $d$, Rogers'
upper bound asymptotically becomes $d \, 2^{-d/2}/e$. Kabatiansky and Levenshtein
\cite{Ka78} found an even stronger bound, which in the limit $d\rightarrow \infty$
yields $\phi_{\mbox{\scriptsize max}} \le 2^{-0.5990d(1+o(1))}$.
Cohn and Elkies \cite{Co03} obtained and computed linear programming upper bounds,
which provided improvement over Rogers' upper bound for dimensions
4 through 36. They also conjectured that their approach could be used
to prove sharp bounds in 8 and 24 dimensions.
Indeed, Cohn and Kumar \cite{Co07} used these techniques
to prove that the Leech lattice is the unique densest lattice in $\mathbb{R}^{24}$. They
also proved that no sphere packing in
$\mathbb{R}^{24}$ can exceed the density of the Leech lattice
by a factor of more than $1+1.65 \times 10^{-30}$,
and gave a new proof that the $E_8$ lattice is
the unique densest lattice in $\mathbb{R}^8$.

A recent investigation \cite{To06a}  proves that there exists a disordered packing
construction in $\mathbb{R}^d$  with a maximal density that achieves the 
saturation lower bound (\ref{sat}) for any $d$. This construction
is referred to as the ``ghost" random sequential addition
(RSA) packing \cite{footnote3} and it was shown that all of the $n$-particle correlation functions 
for this packing can be obtained analytically for all allowable densities and in any dimension.
Interestingly, this packing is {\it unsaturated} (see Fig. 1)  and yet it has a maximal
density $2^{-d}$, which suggests that there exist disordered saturated packings
that exceeds the saturation lower bound (\ref{sat}) or the Minkowski lower bound (\ref{mink}).
Indeed, another recent study \cite{To06c} strongly suggests that the standard
disordered RSA packing \cite{footnote4} at its maximal saturation density
scales as $d \, 2^{-d}$ for large $d$, which has the same asymptotic 
behavior as Ball's lower bound (\ref{ball}). Note that spheres
in both the ghost and standard RSA packings cannot form interparticle
contacts, which appears to be a crucial attribute to 
obtain exponential improvement on Minkowski's bound \cite{To06b},
as we discuss below.

\begin{figure}[bthp]
\centerline{\includegraphics[height=3.5in,keepaspectratio,clip=]{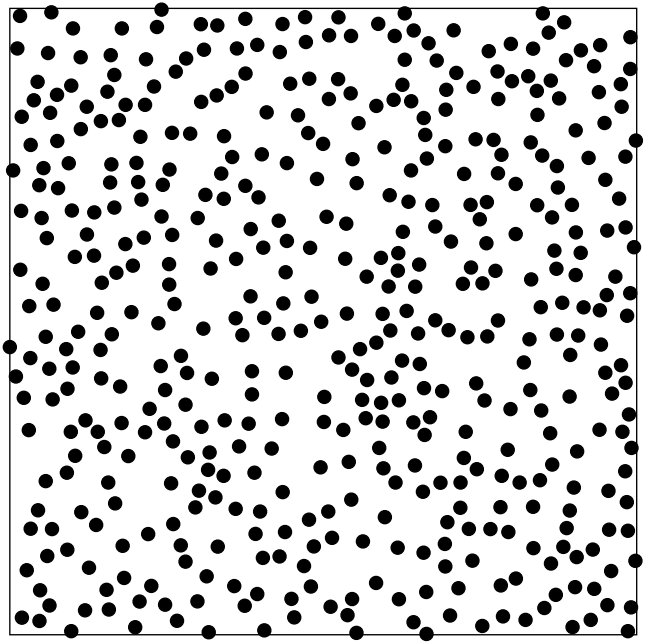}}
\caption{(Color online) A configuration of 468 particles 
of a ghost RSA packing in $\mathbb{R}^2$
at a density very near its maximal density of 0.25. This was generated
using a Monte Carlo procedure within a square fundamental cell under periodic boundary conditions. 
Note that the packing is clearly unsaturated and there are no contacting particles.}
\label{phi-fit}
\end{figure}

Do there exist disordered packings that can provide the long-sought
exponential improvement of Minkowski's lower bound?
Torquato and Stillinger  \cite{To06b} employed a conjecture concerning the existence
of disordered sphere packings  and an optimization procedure that maximizes the density
associated with a ``test" pair correlation function 
to provide the putative exponential improvement on Minkowski's 100-year-old bound
on  $\phi_{\mbox{\scriptsize max}}$ (see Section II for details). The asymptotic behavior
of the conjectural lower bound is controlled by $2^{-((0.77865\ldots)) d}$.
Moreover, this lower bound always lies below the density of the densest
known packings for $3 \le d \le 56$, but, for $d >56$, it can be larger than
the density of the densest known arrangements, all of which are ordered.
These results counterintuitively suggest that the densest packings in sufficiently
high dimensions may be disordered rather than periodic, implying
the existence of disordered classical ground states for some continuous potentials.
In addition, a {\it decorrelation} principle for disordered packings was identified in Ref. \cite{To06b},
which states that {\it unconstrained} correlations in disordered sphere packings
vanish asymptotically in high dimensions and that the $g_n$ for any $n \ge 3$ can be inferred 
entirely (up to some small error) from a knowledge
of the number density $\rho$ and the pair correlation function $g_2({\bf r})$.
This decorrelation principle \cite{footnote5}, among other things, provides
justification for the conjecture used in Ref. \cite{To06b},
and is vividly and explicitly exhibited by the exactly solvable
ghost RSA packing process \cite{To06a} as well as by computer simulations
in high dimensions of the maximally random jammed state \cite{Sk06} and the
standard RSA packing process \cite{To06c}.

In this paper, we investigate whether the putative exponential improvement
of Minkowski's lower bound found in Ref. \cite{To06b}
can be further improved by exploring other test
pair correlation functions. We will show that
there are simpler test functions that lead to
the same asymptotic result. More importantly, we will
demonstrate that there is a wide class of test functions that lead
to the same exponential  improvement as in Ref. \cite{To06b}.

\section{Preliminaries and Optimization Procedure}

A packing of congruent spheres of unit diameter
is simply a point process in which any pair of points cannot be closer
than a unit distance from one another \cite{To06b}.
A particular configuration of a point process 
in $\mathbb{R}^d$ is described by the ``microscopic" density
\beq
n({\bf r})=\sum_{i=1}^\infty\delta({\bf r}-{\bf x}_i).
\eeq
This distribution can be interpreted in a probabilistic sense
\cite{To06b}, which is particularly useful for the
arguments we will present, even in the limit in which no explicit
randomness is present, as in the case in which the spheres are
arranged on the sites of a (Bravais) lattice. We define the $n$-particle density as the
ensemble average 
\beq
\rho_n({\bf r}_1,...,{\bf r}_n)=\avg{\sum_{i_1\neq i_2\neq ...\neq
i_n}\!\!\delta({\bf r}_1-{\bf x}_{i_1})\ ...\
\delta({\bf r}_n-{\bf x}_{i_n})},
\eeq
which is a nonnegative quantity.
Henceforth, we will assume that the random process is translationally invariant,
i.e., statistically homogeneous. It follows  that there is no preferred
origin in the packing and thus the 
$n$-particle densities $\rho_n({\bf r}_{12},{\bf r}_{13},\ldots, {\bf r}_{1n})$ only depend on relative displacements,
where ${\bf r}_{1j}\equiv {\bf r}_j -{\bf r}_1$. In particular, the one-particle density
$\rho_1({\bf r})=\avg{\delta({\bf r}-{\bf x}_1)}=\rho$ is a constant called the {\it number (center) density}.
Note that the packing density $\phi$ defined earlier
is related to the number density $\rho$ for
spheres of unit diameter via the relation
\beq
\phi=\rho v_1(1/2).
\eeq
where $v_1(r)=\pi^{d/2}r^d/\Gamma(d/2+1)$ is the volume of a
sphere of radius $r$. The surface area of such a sphere is
$s_1(r)=2\pi^{d/2} r^{d-1}/\Gamma(d/2)$.
If we divide $\rho_n$ by
$\rho^n$, we get the $n$-particle correlation function $g_n({\bf r}_{12},{\bf r}_{13}...,{\bf r}_{1n})$, which clearly is also a nonnegative function. 
As will become clear shortly, the \emph{pair correlation} function $g_2({\bf r}_{12})$ has
particular importance to us. If the point process is additionally rotationally
invariant (i.e., the packing is statistically homogeneous and isotropic),
the pair correlation function $g_2(r)$ depends only on the
distance $r \equiv |\bf r|$.

In Ref. \cite{To02c}, $g_2$-invariant processes were examined in order
to gain insights about the nature of disordered sphere packings.
A {\it $g_2$-invariant process} is one in which a given nonnegative pair correlation 
$g_2({\bf r})$ function remains invariant for all ${\bf r}$ over the range of
densities
\begin{equation}
0 \le \phi \le \phi_*.
\end{equation}
The terminal density $\phi_*$ is the maximum achievable density
for the $g_2$-invariant process subject to satisfaction of
certain necessary conditions on the pair correlation. 
In particular, they considered those ``test" $g_2(r)$'s that are distributions on $\mathbb{R}^d$ depending
only on the radial distance $r$.
For any test $g_2(r)$ associated with a packing, i.e., $g_2(r)= 0 \; \mbox{for}\;  r<1$, 
they maximized  the corresponding density $\phi$, i.e.,
\begin{equation}
\max\phi
\label{opt}
\end{equation}
subject to the following two conditions:
\beq
g_2(r)\geq 0 \qquad \mbox{for all}\quad  r,
\label{eq:g2geq0}
\eeq
\beq
S(k)=1+\rho(2\pi)^{d/2}\int_0^\infty dr
r^{d-1}\frac{J_{d/2-1}(kr)}{(kr)^{d/2-1}}[g_2(r)-1] \geq 0 \qquad \mbox{for all}\quad  k.
\label{eq:Sgeq0}
\eeq
Condition (\ref{eq:Sgeq0}) states that the structure factor $S(k)$ [trivially related to the Fourier transform
of $g_2(r)-1$] must also be nonnegative for all wavenumbers. It is
a known necessary condition on the existence of a point process \cite{To02c,To03},
but it is generally not sufficient \cite{Cos04}.

Recently, Torquato and Stillinger \cite{To06b} conjectured that 
a disordered sphere packing in $\mathbb{R}^d$ at number density $\rho$
exists for sufficiently large $d$  if and only if
the conditions (\ref{eq:g2geq0}) and (\ref{eq:Sgeq0}) are satisfied. The maximum achievable density
is the terminal density $\phi_*$, which then implies
the lower bound
\begin{equation}
\phi_{\mbox{\scriptsize max}} \ge \phi_*
\end{equation}
There is mounting evidence to support this conjecture.
First, the aforementioned decorrelation principle states that unconstrained
correlations in disordered sphere packings vanish asymptotically in high dimensions
and that the $g_n$ for any $n \ge 3$ can be inferred entirely from a knowledge
of $\rho$ and $g_2$.  Second, other necessary
conditions on $g_2$, such as the Yamada condition \cite{Ya61}
as well as others \cite{To06b}, appear to only have relevance in very low dimensions.
Third, one can recover the form of known rigorous bounds [cf. (\ref{mink}) and (\ref{ball})]
for specific test $g_2$'s when the conjecture is invoked. Finally, in these two instances,
configurations of disordered sphere packings on the torus
have been numerically constructed with such $g_2$ in 
low dimensions for densities up to the terminal density \cite{Cr03,Uc06a}.

Interestingly, the optimization problem defined above is the {\it dual}
of the infinite-dimensional linear program devised by Cohn and Elkies \cite{Co02,Co03}
to obtain upper bounds on the maximal packing density. 
In particular, let $f(r)$ be a radial function in $\mathbb{R}^d$ such that
\begin{eqnarray}
f(r)& \le& 0 \quad \mbox{for}\quad  r \ge 1, \nonumber \\
{\tilde f}(k) &\ge& 0 \quad \mbox{for all} \; k.
\label{cohn-elkies1}
\end{eqnarray}
Then the number density $\rho$ is bounded from above by 
\begin{equation}
\min \frac{f(0)}{2^d{\tilde f}(0)}.
\label{cohn-elkies2} 
\end{equation}
The radial function $f(r)$ can be physically interpreted to be a
{\it pair potential}. The fact that its Fourier transform must be
nonnegative for all $k$ is a well-known stability condition
for many-particle systems with pairwise interactions \cite{Ru99}. We see that whereas
the linear program specified by (\ref{opt}), (\ref{eq:g2geq0}) and (\ref{eq:Sgeq0})
utilizes information about pair correlations, its dual program ({\ref{cohn-elkies1}) 
and ({\ref{cohn-elkies2})  employs information about
pair interactions. It is important to note \cite{To06b} that even if  there does
not exist a sphere packing with $g_2$ satisfying conditions
(\ref{eq:g2geq0}) and (\ref{eq:Sgeq0}), the terminal density $\phi_*$ can never exceed the
Cohn-Elkies upper bound. Every linear program has a dual program and when
an optimal solution exists, there is no {\it duality gap} between
the upper bound and lower bound formulations. However, until recently,
it was not clear how to prove that there was no duality gap for
the aforementioned infinite-dimensional sphere-packing linear program \cite{Co02}.
Recently, Cohn and Kumar \cite{Co07b} have proved that there is no duality
gap.

By means of the linear program described above and the aforementioned  conjecture 
concerning the existence for a certain test function $g_2$, it was found in Ref. \cite{To06b}
that in the limit $d\to\infty$,
\beq
\phi_{\mbox{\scriptsize max}} \ge \phi_*\sim 2^{-\frac{3}{2}d+\frac{1}{\ln
2}\frac{d}{2}+2.12497... d^{1/3}+\frac{1}{6}\log_2 d+\log_2(3.2761...)},
\label{TS}
\eeq
where the terms neglected are monotonically decreasing with $d$.
The first term in the series provides the putative exponential improvement of Minkowski's
lower bound (\ref{mink}). In the following, we will be interested mainly in the exponential 
improvement of Minkowski's lower bound, and so we simplify the right-hand
side of (\ref{TS}) by writing it as
\begin{equation}
\phi_*\sim 2^{-(\frac{3}{2}-\frac{1}{2\ln 2})d}=2^{-0.77865\dots d}.
\end{equation}
This is not to be intended as an asymptotic expansion of $\phi_*$ in the sense of 
Poincar\'e (the ratio of the right-hand side to the left-hand side does not go to 
unity when $d\to\infty$), however, it is an asymptotic expansion in such sense for $\log_2 \phi_*$.

In what follows, we will show that we can obtain a conjectural lower bound 
asymptotically equal to (\ref{TS}) with a simpler test function.
Then  we will demonstrate that the requirement of hyperuniformity \cite{To03} in Ref. \cite{To06b} 
is actually a necessary condition that arises only from the optimization procedure.
Finally, we will show some examples of how enlarging the 
space of test functions where the optimization is performed does not change the asymptotic 
\emph{exponential} behavior, although non-exponential improvement is found.

Although these results do not constitute a proof of lower bounds, they strongly suggest that an 
estimate of the asymptotic behavior of the solutions to the lower-bound linear programming problem can 
be achieved and that physical intuition is gained about the spatial structures they describe.

\section{Step Plus Delta Function Revisited}

Following Torquato and Stillinger \cite{To06b},
we choose the following test $g_2(r)$:
\beq
g_2(r)=\Theta(r-1)+\frac{Z}{s_1(1)\rho}\delta(r-1).
\label{step-delta}
\eeq
Here the parameter $Z$ has the
interpretation of the average kissing number. The structure
factor becomes
\barr
S(k)
&=&1-2^{d/2}\Gamma\left(1+\frac{d}{2}\right)\frac{J_{d/2}(k)}{k^{d/2}}2^{d}\phi+
2^{d/2-1}\Gamma\left(\frac{d}{2}\right)\frac{J_{d/2-1}(k)}{k^{d/2-1}}Z\nonumber\\
&\equiv&1- a(k)\ 2^d\phi+ b(k)\ Z,
\earr
which defines the functions $a,b$.
The terminal density is defined by the linear
program (\ref{opt}), (\ref{eq:g2geq0}) and (\ref{eq:Sgeq0}).
$Z$ is then a free parameter to be optimized appropriately.

Unlike Torquato and Stillinger \cite{To06b}, we do not impose hyperuniformity \cite{To03}
(requiring the structure factor to vanish at $k=0$) to simplify the optimization.
Moreover, we are also interested in finding the 
largest average kissing number $Z$ that (for a given $d$) satisfies the constraints. 
In this latter case, it is $\phi$ that must be chosen appropriately. 
These are two infinite-dimensional, linear programming
problems. 

There is a graphical construction that will help us look for such points and that 
will be helpful also in cases where more parameters are to be varied. For any given 
$k$ the set of allowed points in the $(\phi,Z)$ plane [i.e., those for which $S(k)\geq 0$] is the half plane
above (below) the line $1-a(k)2^d\phi+b(k)Z=0$ for positive (negative) $a$. 
Upon changing k by a small step to $k+\Delta$, we repeat the construction and find the intersection of the two half-planes. By
letting $k$ vary over the positive reals and letting $\Delta\to 0$,
we find a limiting finite, convex region ${\cal B}$ which gives the allowed values of $\phi,Z$. 
This region is the set internal to the curve obtained by solving the equations
\begin{equation}
S(k,\phi,Z)=0,\quad \frac{\partial}{\partial k}S(k,\phi,Z)=0,
\end{equation}
with respect to $\phi,Z$. This is depicted in Fig.\ref{B}. It is not difficult to prove that the region ${\cal B}$ is indeed internal to the entire spiral. It will suffice to observe that the distance of a point on the spiral from the origin is a monotonically increasing function (for sufficiently large $k$).

Now the terminal density $\phi_*$ is the $x$-component of the rightmost point in ${\cal B}$. Analogously the $y$-component of the topmost point in ${\cal B}$ gives the terminal kissing number $Z_{**}$.

\begin{figure}
\begin{center}
\includegraphics[width=11cm]{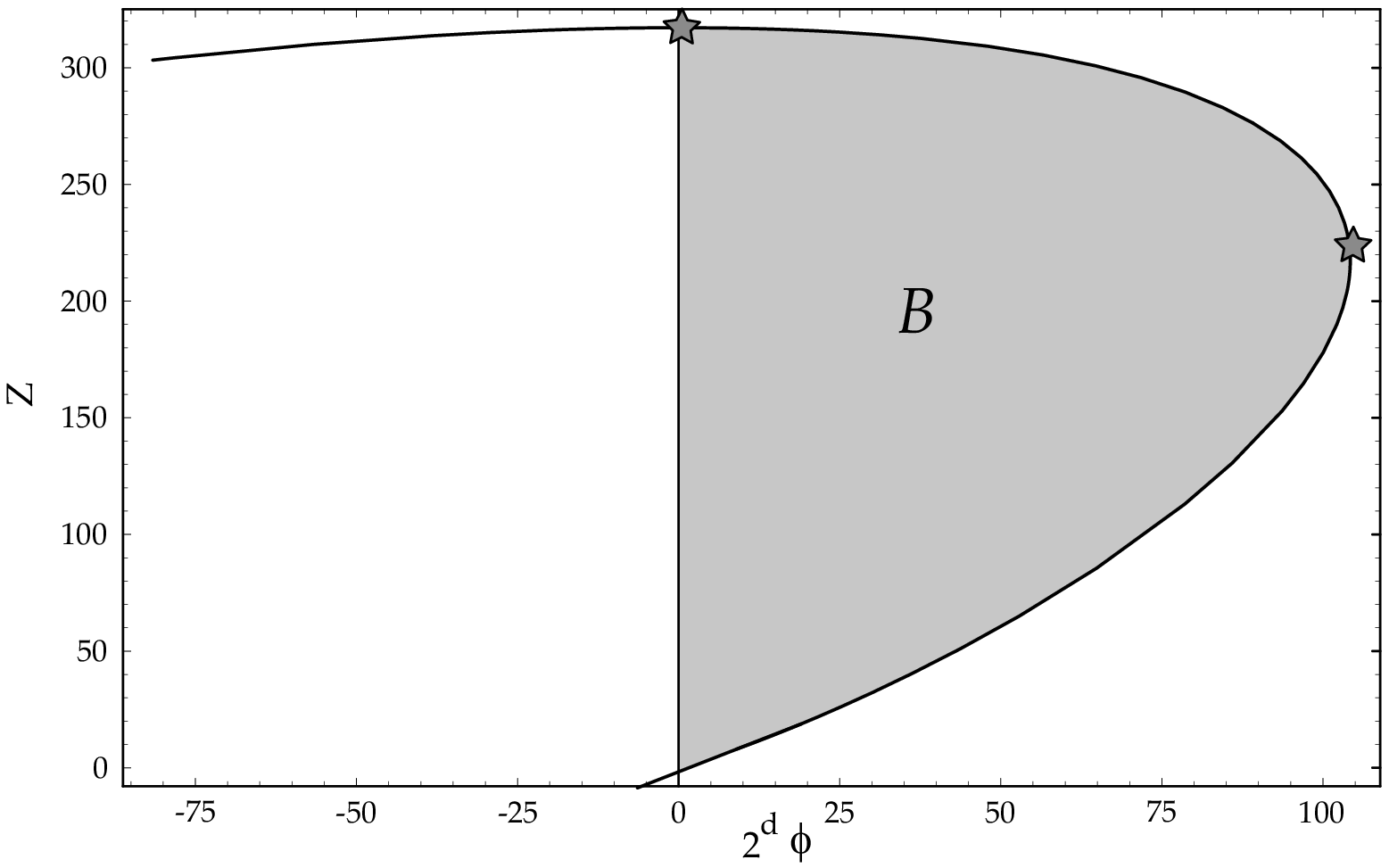}\vspace{1cm}\\
\includegraphics[width=11cm]{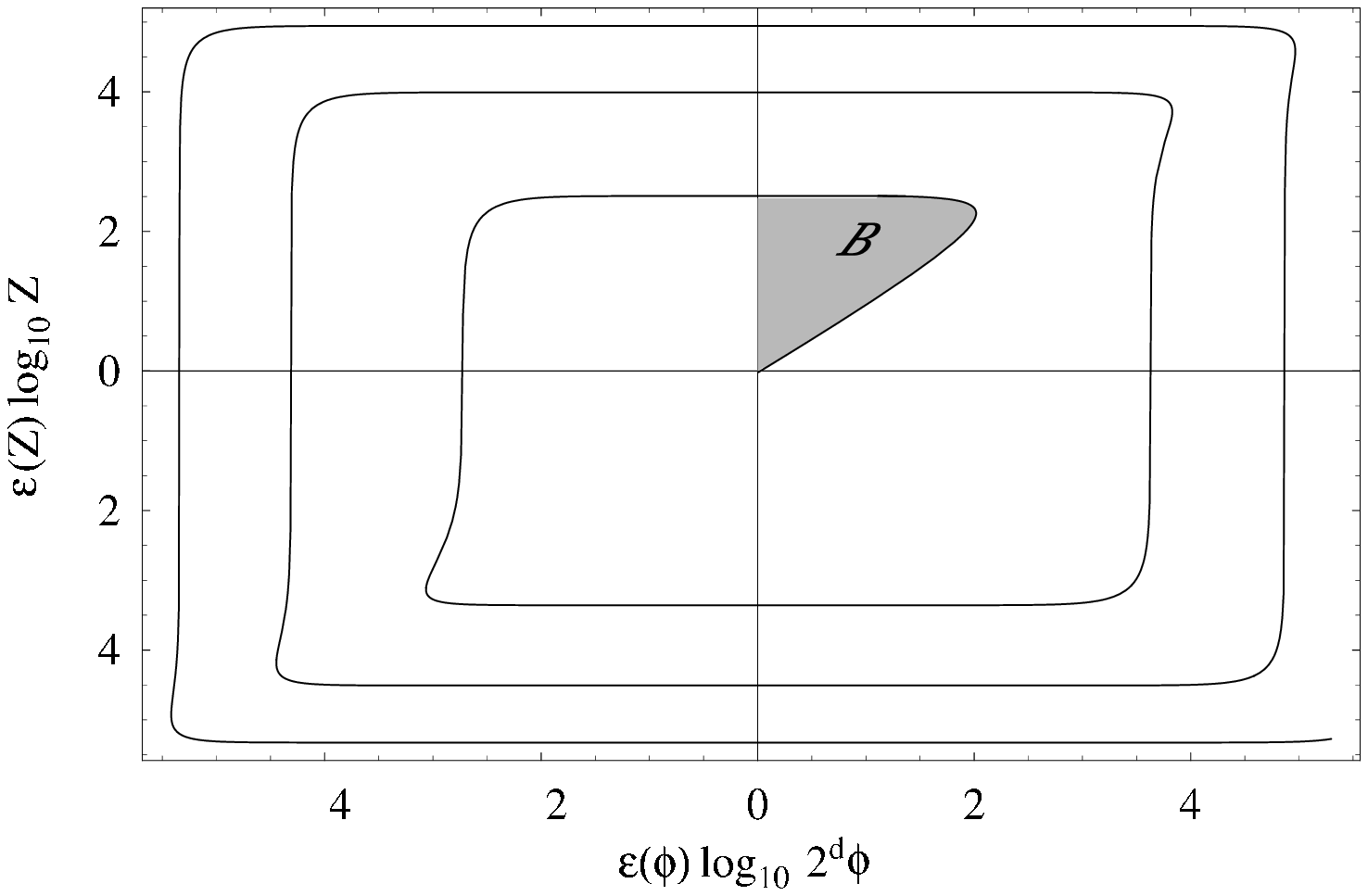}
\caption{\label{B}\small ({\bf Top panel}) For $d=16$, the set ${\cal B}$ of allowed packing densities
and kissing numbers. The rightmost point is the maximal packing density $\phi_*$ and its corresponding kissing number $Z_*$. The topmost point
is the maximal kissing number $Z_{**}$ which corresponds to packing density $\phi_{**}=0$. ({\bf Bottom panel}) As in top panel, the region ${\cal B}$ of allowed packing densities and kissing numbers for $d=16$. 
For convenience in plotting, the horizontal and vertical axes represent the functions $\epsilon(\phi)\log_{10}(|2^d\phi|)$, and $\epsilon(Z)\log_{10}(|Z|)$, where $\epsilon(x)={\rm sign} x$, respectively (although in this way the small
region $|2^d\phi|<1,|Z|<1$ had to be left out of the graph). This figure shows how the solution of the equations $S(k,\phi,Z)=0,\
\partial S(k,\phi,Z)/\partial k=0$ for varying $k$ form an ever-growing spiral in which the allowed region ${\cal B}$ is completely contained. So this geometrical construction proves that every point in ${\cal B}$ are solutions to the linear programming problem $S(k,\phi,Z)\geq 0,\ \phi\geq 0,\ Z\geq 0$ for every $k\geq 0$.}
\end{center}
\end{figure}

The terminal density is found at the first zero of
$b(k)$, which is located at the first zero of the Bessel
function of order $d/2-1$. As customary we call this number $j_{d/2-1,1}$. The value of
$(\phi_*,Z_*)$ is then found by finding the point on the spiral corresponding to $k=j_{d/2,1}$:
\barr
\phi_{*}&=&\frac{2^{-d}}{a(j_{d/2-1,1})}=2^{-3d/2}\frac{(j_{d/2-1,1})^{d/2}}{\Gamma(1+d/2)J_{d/
2}(j_{d/2-1,1})},\\
Z_*&=&\frac{a'(j_{d/2-1,1})}{b'(j_{d/2-1,1})a(j_{d/2-1,1})}=\frac{a'(j_{d/2-1,1})}{b'(j_{d/2-1,1})}
2^d\phi_*.
\earr
By using the asymptotic formulas, valid for large $\nu$
\barr
j_{\nu,1}&=&\nu+1.85576...\ \nu^{1/3}+\Ord{\nu^{-1/3}},\\
J_{\nu}(j_{\nu-1,1})&=&-J'_{\nu-1}(j_{\nu-1,1})=1.11310...\
\nu^{-2/3}+\Ord{\nu^{-4/3}},
\earr
we find
\beq
\phi_*\simeq 2^{-\frac{3}{2}d+\frac{1}{\ln
2}\frac{d}{2}+2.12497...d^{1/3}}\sim 2^{-(0.77865\ldots) d}.
\eeq
Notice that this is the same case that was treated in
\cite{To06b} but there hyperuniformity was imposed and the Minkowski bound was recovered. Here \emph{we are not imposing hyperuniformity}
and the resulting terminal structure factor is not hyperuniform. The form of $S(k)$ at the terminal point $\phi_*,Z_*$ is given in Figure
2. Notice that the first zero is at $k=j_{d/2-1,1}\simeq d/2$. This can be
interpreted as the appearance of a structure with length-scale
${\ell}\sim 1/d$ in the system at large $d$. However, since a sphere packing
corresponding to such an $S(k)$ could not be hyperuniform,
it cannot be a Bravais lattice. 

Following \cite{To06b}, we check whether the Yamada condition \cite{Ya61} on the 
number variance \cite{Ya61} is satisfied by the pair correlation (\ref{step-delta}).  
As in \cite{To06b}, we find a violation only for $d=1$.

\begin{figure}
\centerline{\includegraphics[width=10cm]{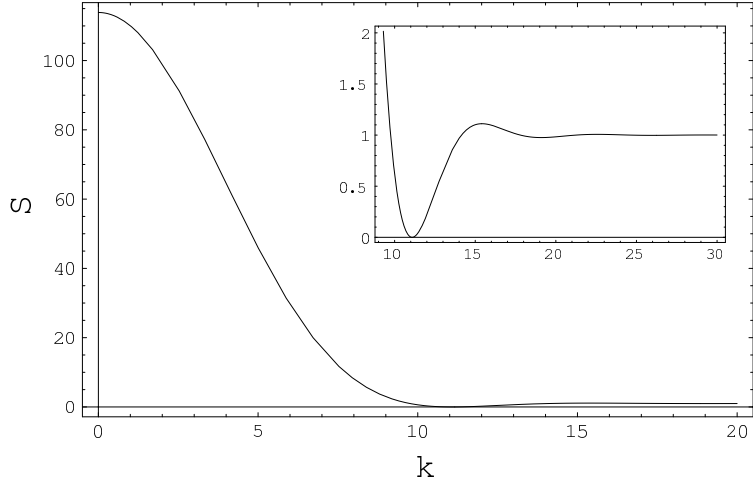}}
\caption{The structure factor for the terminal density $\phi_*=0.0046692,\ Z_*=217.121$ in $d=16$. 
Notice the zero at $k^*=j_{7,1}=11.086...$ and the rapid asymptoting to the value $S(\infty)=1$.}
\end{figure}

The terminal kissing number is given by the topmost point
in ${\cal B}$ which is the point $k^{**}$ where $a(k^{**})=0$. It
can be easily proved that $b'(k^{**})=0$ as well so that
$\phi_{**}=0$ and
\beq
Z_{**}=-\frac{1}{b(j_{d/2,1})}\sim 2^{(\frac{1}{\ln
2}-1)\frac{d}{2}}.
\eeq
It is intriguing to notice that the density corresponding to the terminal kissing number is zero.

\section{Step Plus Delta Function with a Gap}

This case was analyzed by \cite{To06b} before by imposing
hyperuniformity. Here we show that in order to find the terminal density, one does not need 
to impose hyperuniformity from the beginning but rather that it arises as a
necessary condition form the optimization procedure. We will show that the same asymptotic behavior of the 
terminal density found in the previous example  is obtained (modulo non-exponential prefactors). 

We choose the test function
\beq
g_2(r)=\Theta(r-(1+\sigma))+\frac{Z}{s_1(1)\rho}\delta(r-1),
\eeq
depending on two parameters, $Z,\sigma$ and the density of centers $\rho$. Performing the integrals gives the corresponding structure factor
\beq
S(k)=1-a((1+\sigma)k) 2^d(1+\sigma)^{d}\phi+ b(k) Z,
\eeq
where the functions $a,b$ were defined in the previous section.
Again we look for the rightmost point in the set, which is now
given by
\barr
\phi_*&=&\frac{2^{-d}}{(1+\sigma)^{d}a((1+\sigma)j_{d/2-1,1})}\\
Z_*&=&\frac{(1+\sigma)a'((1+\sigma)k)}{b'(j_{d/2-1,1})a((1+\sigma)j_{d/2-1,1})}.
\earr
We now need to maximize the value of $\phi_*$ over $\sigma$. Clearly, we 
can increase $\sigma$ to increase $\phi_*$ indefinitely until
$a((1+\sigma)j_{d/2-1,1})$ becomes zero, namely when
$(1+\sigma)j_{d/2-1,1}=j_{d/2,1}$, which gives $\sigma\sim 2/d$.
The prefactor goes to a constant: $(1+\sigma)^{d}\sim (1+2/d)^d\sim e^{2}$
and does not change the asymptotic dependence on $d$. This would suggest that the density 
can be increased without bound by adjusting the other parameters. This is not the case, however, 
since when we increase $\sigma$ we encounter the first ``global" obstacle [by which 
we mean at wavenumbers $k$ far from the first zero of $b(k)$, which was setting the relevant $k$ scales up to now]
at the value of $\sigma$ when $(1+\sigma)^{d}2^d\phi_*=Z_*-1$.
Notice that $a(0)=b(0)=1$ and both functions decrease
monotonically until their first zeros; here we have
$S(0)=1-(1+\sigma)^{d}2^{d}\phi_*+Z_*=0$ and any further increase of $\sigma$ would make $S(0)<0$. 
\emph{Thus, hyperuniformity has arisen as an optimality condition}. 
Of course one should make sure that there is not a
disconnected region in the parameter space $(\sigma,\phi,Z)$ with better terminal density $\phi_*$ 
but where hyperuniformity does not hold. We have searched the parameter space by discretizing the relevant range of $k$
and solving, using Mathematica, the linear program (\ref{opt}), (\ref{eq:g2geq0}) and (\ref{eq:Sgeq0}). We have
not been able to find another allowed region of the parameters disconnected from the previous one.

Hence we assume that the global terminal value $\phi_*$ is indeed obtained by imposing hyperuniformity and 
maximizing with respect to the remaining parameters (the two operations can be performed in any order). 

We notice that now we have reduced the problem to the case that has been analyzed in \cite{To06b}. We will not 
repeat that analysis here but refer the reader to that paper. It is important to observe that in 
\cite{To06b} the resultant asymptotic scaling law for the terminal fraction $\phi_*$
coincides with the one presented in the previous section $\phi_* \sim 2^{-(0.77865\ldots) d}$. Although the 
non-exponential terms are different from those in the previous section, it is remarkable
that the same exponential scaling law arises for two different cases. This strongly suggests that a 
large class of test functions can possess this asymptotic behavior. With this in mind, we 
go on to analyze the next case win which the test pair correlation function 
consists of a hard core with two delta functions and a gap.  

\section{Step Plus Two Delta Functions with a Gap}

In this Section we find the solution of the optimization problem (\ref{opt}), (\ref{eq:g2geq0}) and (\ref{eq:Sgeq0})
for the family of pair correlation functions $g_2(r)$ 
composed of  unit step function plus a gap and two delta functions, one at contact and the other at the end of the gap:
\begin{equation}
g_2(r)=\theta(r-(1+\sigma))+\frac{Z_2}{s(1)\rho}\delta(r-1)+\frac{Z_1}{s(1+\sigma)\rho}\delta(r-(1+\sigma)).
\end{equation}
This family depends on 3 parameters, $\sigma,Z_1,Z_2$ and we need to optimize them in order to find 
the optimal terminal density $\phi_*$. The structure factor is
\begin{eqnarray}
S(k)&=&1+Z_2 2^{d/2-1}{\Gamma(d/2)}\frac{J_{d/2-1}(k)}{k^{d/2-1}}+Z_12^{d/2-1}\Gamma(d/2)\frac{J_{d/2-1}(k(1+\sigma))}{(k(1+\sigma))^{d/2-1}}+\nonumber\\ 
&-&\phi\Gamma(d/2+1)(1+\sigma)^{d} 2^{3d/2}\frac{J_{d/2}((1+\sigma) k)}{(k(1+\sigma))^{d/2}}\\
&\equiv&1+Z_2\ c(k)+Z_1\ b(k)-(1+\sigma)^{d} 2^d\phi\ a(k),
\label{eq:S2del-gap}
\end{eqnarray}
where the last line defines the functions $a,b,c$. Notice that $a(0)=b(0)=c(0)=1$ and $|a(k)|,|b(k)|,|c(k)|\leq1$ follow 
from the properties of the Bessel functions. It is also convenient to reabsorb the factor $(1+\sigma)^{d} 2^d$ 
in the definition of $\phi$, i.e.\ $(1+\sigma)^{d} 2^d\phi\to\phi$. We will restore the proper units at the 
end of the calculation. The solution of this optimization problem for arbitrary $d$ is a formidable task. 
However, guided by the results of the previous section, we assume we can find an improvement on the 
previous bound even after imposing hyperuniformity. 

Therefore, we fix the value of $Z_2=\phi-Z_1-1$ and are left with the other two 
parameters to optimize. Inserting this value of $Z_2$ in (\ref{eq:S2del-gap}),
we find the reduced optimization problem
\begin{equation}
S(k)=(1-c(k))-(a(k)-c(k))\phi+(b(k)-c(k))Z_1\geq 0.
\end{equation}
By using the fact that $c(k)\leq 1$ we might as well study the optimization problem 
\begin{eqnarray}
S^{(1)}(k,\sigma,\phi,Z_1)&\equiv& \frac{S(k)}{1-c(k)}\equiv 1-\alpha(k)\phi+\beta(k)Z_1\geq 0,\\
\alpha(k)&=&\frac{a(k)-c(k)}{1-c(k)},\\
\beta(k)&=&\frac{b(k)-c(k)}{1-c(k)}.
\end{eqnarray}
Formally, this problem is analogous to the previous case with one delta function with gap and 
can be studied in the very same fashion. The process of having solved for $Z_2$ and changed 
the functions $a,b$ to $\alpha,\beta$ can be thought of as a \emph{renormalization} process 
that allows to \emph{integrate out} one delta function to reduce the problem to a simpler one. 

The mathematical problem of finding the terminal fraction is formally identical to that of the previous 
section, although the constitutive functions $\alpha,\beta$ are more complicated. However, as long as 
a numerical analysis is concerned this does not present further difficulties. 

We proceed in the following way: for a fixed $\sigma$ we find the rightmost point of allowed region, 
$\phi_*(\sigma),\ Z_{1,*}(\sigma)$, by finding the first zero of $\beta(k)$, call it $k^*$,  
\begin{eqnarray}
\phi_*(\sigma)&=&\frac{1}{\alpha(k^*)},\\
Z_{1,*}(\sigma)&=&\frac{\alpha'(k^*)}{\beta'(k^*)\alpha(k^*)}\ .
\end{eqnarray}

We then maximize the value of $\phi_*(\sigma)$ with respect to variations of $\sigma$. Generically, 
increasing $\sigma$ increases the value of $\phi_*$ until a positivity 
condition is violated (for small $k$). It turns out that the first condition to be 
violated is $S^{(1)}(0)\geq 0$. So in practice we find the terminal value of $\sigma$ by solving the equation
\begin{equation}
S^{(1)}(0,\sigma,\phi_*(\sigma),Z_{1,*}(\sigma))=0,
\end{equation}
with respect to $\sigma$. Notice that this is now a ``strong'' hyperuniformity requirement, 
since $S^{(1)}(k)\sim k^2$ near the origin implies $S(k)\sim k^4$ near the origin,
since $1-c(k)\sim k^2$. We are tempted to conjecture that this is a universal feature: 
\emph{adding more delta-functions to $g_2$ and solving the linear programming problem, 
we obtain structure factors $S(k)$ that become increasingly flatter at the origin}. 
Hence, at least in this respect,  the structure factor looks 
increasingly similar to that of a lattice.

As can be seen from Table 1 and Figure \ref{fig:comparison} (here the proper normalization for $\phi$ has been restored) 
the improvement on the previous bound is relevant \emph{but the asymptotic exponent is the same}. Analytically, it is not 
difficult to obtain the rate of exponential decay (dictated mainly by the Stirling expansion of the gamma functions 
and the scaling of the first zero of $\beta$ with $d$ for large $d$),
which turns out to be the {\it same} as the previous cases, namely,
\begin{equation}
\phi_{*}\sim 2^{-(3/2-1/2\ln 2) d}.
\label{exp}
\end{equation}
It is plausible, therefore, that the incorporation of any finite number of delta functions
in a test $g_2$ will not improve the exponent in (\ref{exp}). 
This exponent fits the numerical data very well. A best fit of the data in Table 
\ref{tb:numbers2} using the functions $d, d^{1/3},\log_2 d$, appearing in the analysis 
in the previous section and invoking the existence conjecture of Ref. \cite{To06b}
yields the putative lower bound
\begin{equation}
\phi_{\mbox{\scriptsize max}} \ge \phi_{*}\simeq 2^{-(0.77865\ldots) d +2.12(\pm 0.04)d^{1/3}+0.39(\pm0.08)\log_2(d)+...}.
\end{equation}
The first term is fixed by our analysis, the $d^{1/3}$ is term is consistent with the analytic 
value 2.1247 in Eq.\ (\ref{TS}). The sub-leading term $\log_2 d$ in this expression is 
very difficult to obtain analytically and we have not succeeded in this task. However, it is 
clear that there there is an improvement from the value $\frac{1}{6}=0.1666\ldots$ 
appearing in (\ref{TS}). The improvement is also evident from the numbers in Table \ref{tb:numbers1}.

It is worth noting that for large $d$ the optimum gap $\sigma\simeq \frac{2.77\ldots}{d}$ (from a best fit analysis). 
This scaling with $d$ is slightly different from that found in the previous section and in \cite{To06b} 
(there $\sigma\simeq 1.81/d$). Again notice that the scaling of $\sigma$ with $d$, $\sigma\propto 1/d$ is necessary 
in order not to introduce an exponential suppression of density. In fact for large $d$, $(1+c/d)^{d}\to e^{c}$ 
multiplies the density $\phi$ in all the formulas (and hence it reduces the terminal value by $e^{-c}$). 
A larger gap, say $\mathcal{O}(d^{-(1-\epsilon)})$, would suppress the density by an exponentially large 
amount $e^{-d^\epsilon}$. 

Table \ref{tb:numbers1} compares the final results of our analysis for the conjectured 
lower bound on the maximal density to the previous lower bound, 
the best known packings, and the optimal upper bound in \cite{Co03} 
for selected dimensions up to $d=150$.
As in the previous cases, the Yamada condition \cite{Ya61} is violated only for $d=1$.
This supports the conclusion reached in Ref. \cite{To06b}
that the Yamada condition appears to only have practical relevance in very
low dimensions. 

\begin{table}[htdp]
\caption{Estimates of the maximal densities for selected dimensions up
to $d=150$. $\phi_{b.k}$ is the densest known packing, $\phi_{CE}$ is the upper bound of Cohn and Elkies, $\phi_{1,*}$ is the terminal density for a single delta function and $\phi_{2,*}$ for two delta functions.}

\centering
\begin{tabular}{c|c|c|c|c}

$d$ & $\phi_{b.k.}$ & $\phi_{CE}$ & $\phi_{*,1}$ & $\phi_{*,2}$  \\
\hline
3 & 0.74049 & 0.77982 & 0.57665 & 0.63306 \\
4 & 0.61685 & 0.64774 & 0.42526 & 0.47885 \\
5 & 0.46527 & 0.52506 & 0.30591 & 0.35437 \\
6 & 0.37295 & 0.41776 & 0.21360 & 0.24966 \\
7 & 0.29530 & 0.32757 & 0.14713 & 0.17991 \\
8 & 0.25367 & 0.25367 & 0.09985 & 0.12467 \\
12& 0.04945 & 0.08384  & 0.01915 & 0.025721   \\
15& 0.01685 & 0.03433 & 0.00516 & 0.00722 \\
19& 0.004121 &    0.009885 &   0.000845 &    0.001233    \\
24 & 0.00193 & 0.00193 & $8.24\times 10^{-5}$ & 0.000125 \\
31&   $1.18\times 10^{-5}$    &   $1.93\times 10^{-4}$      &  $2.91\times 10^{-6}$ & $4.57 \times 10^{-6}$ \\  
36 & $6.14\times 10^{-7}$ & $3.59\times 10^{-5}$ & $2.57\times 10^{-7}$ & $4.13\times 10^{-7}$ \\ 
56 &$2.33\times 10^{-11} $& ---  & $1.25\times 10^{-11}$ &$2.13\times 10^{-11}$ \\
60 &$2.97\times 10^{-13}$& --- & $1.67\times 10^{-12}$ & $2.87\times 10^{-12}$\\
64 & $1.33\times 10^{-13}$& --- & $2.22\times 10^{-13}$ & $3.83\times 10^{-13}$\\ 
80 & $1.12\times 10^{-16}$& --- & $6.52\times 10^{-17}$ & $1.15\times 10^{-16}$ \\
100 &  --- & --- & $2.28\times 10^{-21}$ & $4.11\times 10^{-21}$\\
150 & $ 8.44\times 10^{-39}$ & --- & $1.27\times 10^{-32}$ & $2.30\times 10^{-32}$\\

\end{tabular}
\label{tb:numbers1}
\end{table}

\begin{figure}
\centerline{\includegraphics[width=12cm,keepaspectratio,clip=]{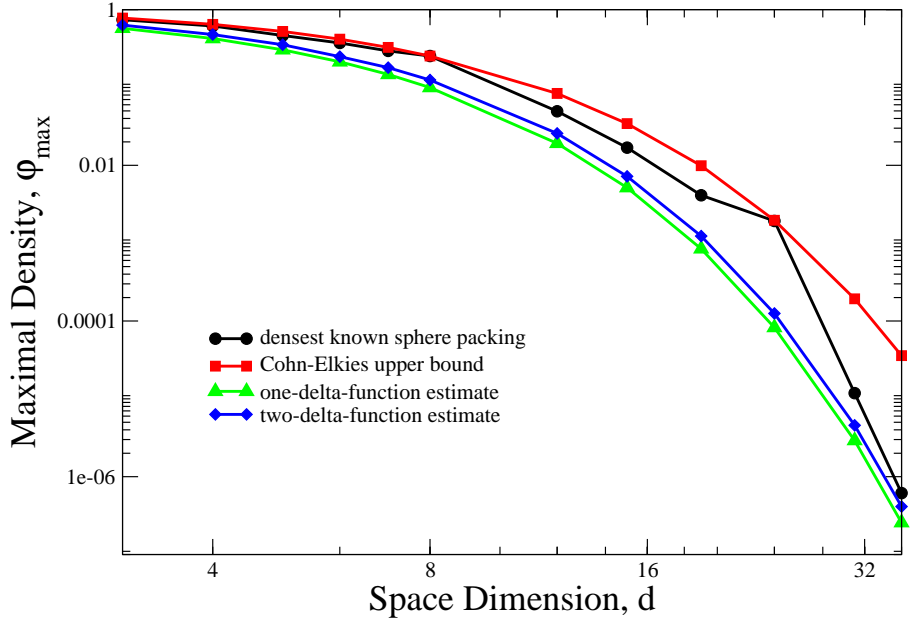}}
\caption{(Color online) \small Comparison of different results for 
the maximal density $\phi_{\mbox{\scriptsize max}}$ versus dimension $d$. From bottom to top: 
Torquato-Stillinger result \cite{To06b} (1-delta function with gap), one
of the results of this paper
(2-delta functions with a gap), densest known packings \cite{Co93},
and the  Cohn-Elkies upper bound \cite{Co03}. \label{fig:comparison}}
\end{figure}

\begin{table}[htdp]
\caption{Terminal density $\phi_{*}$ for two delta functions and a gap, 
corresponding optimal gap $\sigma$, and optimal average kissing number $Z_{1,*}$ for large $d$.}
\centering
\begin{tabular}{c|c|c|c}

$d$ & $\sigma$ & $Z_{1,*}$ & $\phi_{*}$ \\
\hline
200 & 0.013508 & $1.57\times 10^{18}$ & $1.06 \times 10^{-43}$ \\
250 & 0.010895 & $7.15\times 10^{21}$ & $4.18\times 10^{-55}$ \\
300 & 0.009132 & $2.94\times 10^{25}$ & $1.49\times 10^{-66}$ \\
350 & 0.007862 & $1.12\times 10^{29}$ & $4.96\times 10^{-78}$ \\
400 & 0.006903 & $2.93\times 10^{25}$ & $1.56\times 10^{-89}$ \\
450 & 0.006154 & $1.38\times 10^{36}$ & $4.73\times 10^{-101}$ \\
500 & 0.005553 & $4.67\times 10^{39}$ & $1.40\times 10^{-112}$ \\

\end{tabular}
\label{tb:numbers2}
\end{table}

\section{Conclusions and Open Questions}

The problem of finding the asymptotic behavior of the maximal density $\phi_{\mbox{\scriptsize max}}$
of sphere packings in high dimensions is one of the most fascinating and challenging problems in geometry. In this 
paper, we have shown how, using linear programming bounds and a conjecture 
concerning the existence of disordered sphere packings based on pair-correlation information, 
the asymptotic conjectural lower bound \cite{To06b}
\begin{equation}
\phi_{\mbox{\scriptsize max}} \ge 2^{-(0.77865\ldots)d},
\label{asymp}
\end{equation}
which  provides the putative exponential improvement on
Minkowski's century-old lower bound (\ref{mink}), is actually much 
more general than one could have initially surmised. 
Precisely the same exponential improvement arises for a simpler pair-correlation function than the 
one employed in \cite{To06b} and survives also to a considerable enlargement of the family of 
test functions $g_2$. This family  of functions includes two delta functions with
a gap (which we have shown improves upon the prefactor multiplying $2^{-(0.77865\ldots)d}$ given
in Ref. \cite{To06b}) and, we argue, any finite number of 
delta functions. If this is true, as we believe, it signifies that the \emph{decorrelation principle} 
alone has a huge predictive power, since an exponential improvement of Minkowski's bound has 
proved to be an extremely difficult problem. 

One outstanding open question is certainly in which sense this is to be interpreted as an 
asymptotic bound. Based on our present, limited knowledge of optimal sphere packings, we foresee 
diverse scenarios. In one case, for sufficiently large $d$, the importance of higher-order 
correlations is to be neglected altogether and the bound becomes exact by virtue of the 
\emph{decorrelation principle}. This would mean
that the asymptotic Kabatiansky-Levenshtein upper bound is far from
optimal: a provocative possibility. In a second scenario, it could be that ``special dimensions'' 
continue to exist for which the negligence of higher-order correlations is impossible. In this case, 
the lower bound obtained by our methods would not apply to these special dimensions but will 
continue to apply to the other dimensions. On the other hand, if the frequency of appearance of 
these dimensions over the integers is decreasing then the decorrelation principle is safe. 
A third but more pessimistic possibility is that these dimensions 
are actually becoming more and more frequent, and our conjectural bound would apply only to 
the subset of dimensions remaining. However, there is absolutely no evidence at present for 
either the second or third scenario. Our best guess at the moment
is that the optimal packings in very high dimensions
will possess no symmetry at all and therefore are truly disordered.
If so, then the decorrelation principle dictates that pair correlations
alone completely characterize the packing in high $d$, implying that the
form of the asymptotic bound (\ref{asymp}) is exact! 

The fact that pair correlations can completely specify an optimal packing may
seem to be counterintuitive at first glance, but we can now identify even low dimensions
where this phenomenon occurs. 
Specifically, whenever the linear programming bounds are exact (i.e., achieve 
some packing), pair correlation information is sufficient
to determine the optimal packing! This outcome, in all likelihood, occurs in 
$\mathbb{R}^2$, $\mathbb{R}^8$ and $\mathbb{R}^{24}$ \cite{Co03,Co07}.
This implies that whenever linear programming bounds are not sharp
in low dimensions (albeit without a duality gap for any $d$ \cite{Co07b}), 
information about high-order correlations are required to get optimal solutions.

Another interesting question arises because our procedure, like Minkowski's, is nonconstructive.
Specifically,  it is an open question
whether there exist packing constructions that realize our test $g_2$'s.
For future investigations, it would be fruitful  to determine whether there
are periodic or  truly disordered packings that  have pair correlation functions that approximate well the ones 
studied in this paper. If these packings could be identified, one should  attempt
to ascertain whether the higher-order correlations diminish in importance
as $d \to\infty$  in accordance with the decorrelation principle. If such packings
exist (or better, if a $d$-dependent family of them does), they would enable 
one to place on firm, solid ground the putative exponential improvement on Minkowski's bound.
We are currently investigating these questions.

\begin{acknowledgments}
We thank Henry Cohn and Abhinav Kumar for discussions and for making us aware of their 
unpublished  proof that there is no duality gap in the linear programming bounds.
This work was supported by the Division of Mathematical Sciences at
the National Science Foundation under Grant No. DMS-0312067.
\end{acknowledgments}


\begin{thebibliography}{20}
\bibitem{footnote2}
Already in $\mathbb{R}^{10}$, we encounter for the first time
a non-lattice (i.e., periodic) packing that is denser than
all known lattice packings; see Ref. \cite{Co93}.
\bibitem{footnote4}
The well-known standard RSA packing is also  a time-dependent packing
in $\mathbb{R}^d$  in which there are no spheres at 
time $t=0$. For any $t >0$, sphere placements 
are attempted by a purely random process
and are accepted if they do not overlap 
any current sphere in the packing.
In the infinite-time limit, the packing is
saturated and its maximal density is achieved.
\bibitem{footnote5}
Based upon our knowledge of spin systems in high dimensions,
it may be tempting to conclude that the decorrelation principle
is an expected  ``mean-field" behavior, but this cannot be the case
for reasons that have already been discussed elsewhere \cite{Sk06,To06c}.
\end{thebibliography}
\end{document}